# Micromagnetic study of a spin-torque oscillator based on a magnetic nano-contact magnetized at an arbitrary angle

G. Consolo[1*], B. Azzerboni[1],

[1] *University of Messina, Dipartimento di Fisica della Materia e Tecnologie Fisiche Avanzate, Salita Sperone 31, Vill.S.Agata, Messina 98166, ITALY*

L. Lopez-Diaz[2],

[2] *University of Salamanca, Departamento de Fisica Aplicada, Plaza de la Merced s/n, Salamanca 37008, SPAIN*

G. Gerhart[3], E. Bankowski[3],

[3] *U. S. Army TARDEC, Warren, Michigan 48397, USA*

V. Tiberkevich[4], and A.N. Slavin[4]

[4] *Oakland University, Departament of Physics, Rochester, Michigan 48309, USA*

(Dated: 5 March 2008)

**ABSTRACT**

The nature of spin wave modes excited by spin-polarized direct current in a spin-torque auto-oscillator based on a magnetic nanocontact was studied by a micromagnetic simulation in the case when the external bias magnetic field was rotated from the in-plane to perpendicular-to-plane orientation. In qualitative agreement with the weakly-nonlinear analytical theory it was found, that at a certain critical angle, an abrupt switching from the self-localized nonlinear "bullet' mode to a propagating quasi-linear Slonczewski mode takes place, and is accompanied by an upward jump in generated microwave frequency. It was, also, found that the analytical theory overestimates the magnitude of a critical magnetization angle, corresponding to the mode switching, and that the magnitude of the frequency jump caused by the mode switching is inversely proportional to the nanocontact radius.

## INTRODUCTION

Recently, it was demonstrated both theoretically[1-3] and experimentally[4-8] that when a direct electric current traverses a magnetized multilayered magnetic structure it becomes spin-polarized and transfers spin angular momentum between the magnetic layers. This transfer can induce a persistent microwave magnetization precession in the thin ("free") magnetic layer of the structure. The underlying physical mechanism responsible for the current-induced microwave generation is the compensation of the natural positive magnetic damping (which is caused in metallic magnets, for the most part, by the spin-electron interaction), by the current-induced negative effective damping[3]. This effect opens a possibility to develop high-quality microwave spin-torque nano-oscillators, controllable both by the bias magnetic field and by the bias current[4-8]. The practical design of spin-torque nano-oscillators requires deep understanding of the spatial structure and properties of the microwave spin wave modes excited by the direct current. This is especially true for the case of so-called *nanocontact* geometries[5-8], where the lateral sizes of the free magnetic layer are so large that it can be treated as an infinite plane, and the spin-polarized current flows only within a small part (nanocontact area) of the free layer. In this case, there are no lateral boundaries that can determine the spatial structure of the excited spin wave mode, and the determination of the structure of the excited spin wave mode requires a detailed investigation.

The experimental and theoretical studies of spin-torque oscillators based on the nanocontact geometry and performed for different orientations ($\theta_{ext}$) of the external bias magnetic field ($\mathbf{H}_{ext}$) have demonstrated qualitatively different pictures of excited spin wave modes for different external field orientations[5-10]. In the case of a normally magnetized film ($\theta_{ext} = 90°$) both linear[3] and nonlinear[9] theories based on the small-amplitude expansion of the Landau-Lifshitz-Gilbert-Slonczewski (LLGS) equation for the magnetization of the free layer predict that the spin wave mode excited by spin-polarized current is an exchange-dominated propagating cylindrical spin wave with the wave vector inversely proportional to the nanocontact radius and the frequency, that is higher than the frequency of a linear ferromagnetic resonance (FMR). In contrast, in the case of an in-plane magnetized ($\theta_{ext} = 0$) nanocontact, it was shown in Ref.10 that the balance between the dispersion and nonlinearity in the magnetic "free" layer of a nanocontact leads to the formation of an evanescent self-localized standing nonlinear spin wave "bullet" mode,[10] having imaginary wavevector and the frequency that is lower than the linear FMR frequency. The existence of these rather different scenarios of current-induced spin wave excitations was later confirmed in micromagnetic simulations[11-14]. In particular, the numerical investigations performed in Ref. 14 allowed us to confirm with certainty the "subcritically-unstable"[15] nature of the spin wave "bullet" mode excited in the case of in-plane magnetization.

The results of experimental investigations of current-induced microwave excitation in obliquely magnetized nanocontacts were reported in Ref. 8. In that work the authors observed at some values of the bias current the existence of multiple peaks, corresponding to generation of several microwave frequencies that are not harmonics of each other. The authors of Ref. 8, also, observed non-monotonic behavior of the frequency of some of the generated peaks as a function of the bias current and abrupt jumps in the values of generated frequency at certain bias current values.

Although the exact nature of all the spin wave modes observed in the experiment[8] is still not completely clear, the analytic results presented in Ref. 16 suggest that the origin of the observed abrupt frequency jumps is related to the mode-hopping between the quasi-linear propagating mode[3] and the nonlinear evanescent "bullet" mode.[10]

It was demonstrated in Ref. 16, that in the case of an in-plane magnetized nano-contact a nonlinear evanescent "bullet" mode can coexist with the propagating quasi-linear exchange-dominated spin wave mode, but the threshold current $I_{th}^B$ corresponding to the excitation of a "bullet" is substantially lower than the threshold current $I_{th}^L$ corresponding to the excitation of a quasi-linear propagating spin wave mode.

The approximate analytic theory[16] has, also, demonstrated that when the direction of the external magnetic field is tilted from the in-plane towards the perpendicular-to-plane orientation the nonlinear "bullet" mode exhibits an excitation threshold smaller than the one of the quasi-linear propagating mode for magnetization angles $\theta_{ext}$ up to a certain critical angle $\theta_{cr}$ ($I_{th}^B < I_{th}^L$, for $\theta_{ext} < \theta_{cr}$), while the opposite case occurs for the larger magnetization angles ($I_{th}^L < I_{th}^B$, for $\theta_{ext} > \theta_{cr}$). The theoretical dependence of the threshold current for the spin wave excitation on the magnetization angle obtained in Ref. 16 is a continuous function, which has a kink at the transition between the two excited modes (*i.e.*, at $\theta_{ext} = \theta_{cr}$). Consequently, at the point of mode-switching the generated frequency experiences an abrupt frequency jump of the order of several GHz.

It is worth noting that, as it was shown in Ref. 17, the nonlinear frequency shift coefficient $N$ is negative for the in-plane magnetized nanocontact ($\theta_{ext} = 0$) and positive in the case of perpendicular-to-plane magnetization ($\theta_{ext} = 90°$). Thus, there exists a second critical angle ("linear" angle) $\theta_{lin}$, at which the nonlinear frequency shift vanishes ($N = 0$), and above which (i.e. for $\theta_{ext} > \theta_{lin}$) the evanescent spin wave "bullet" mode cannot exist at all.

Although the analytic theory presented in Ref. 16 gives a qualitative picture of spin wave mode excitation by spin-polarized current and mode hopping when the magnetization angle is varied, this approximate theory based on a weakly-nonlinear approach does not allow one to perform quantitative analysis of spin wave excitation above the generation threshold, when the amplitudes of the excited spin waves are too large to be treated perturbatively. Thus, the main goal of our current paper is to present the results of micromagnetic simulations of current-driven spin wave excitation in a magnetic nanocontact magnetized at an arbitrary out-of-plane angle (0° < $\theta_{ext}$ < 90°), and to elucidate the limits of applicability of the approximate analytic theory to a real laboratory experiment. Depending on the value of the magnetization angle $\theta_{ext}$, a number of different hysteretic and non-hysteretic scenarios of spin wave excitations were found in our micromagnetic simulations. The main lesson learned from this micromagnetic numerical experiment is that, although the analytic picture of abrupt frequency jumps caused by dynamic hopping between the quasi-linear and "bullet" modes given by the approximate theory is, in general, qualitatively correct, the actual values of critical angles for these jumps could be very different from those predicted analytically.

## NUMERICAL MODELING

We consider a layered magnetic structure consisting of a thick magnetic layer, called "pinned layer" (PL), a thin non-magnetic spacer, and a thin magnetic layer, called "free layer" (FL), as shown in Fig. 1. A static external magnetic field $\mathbf{H}_{ext}$ and a perpendicular-to-plane direct current $I$ are simultaneously applied to the above described layered structure. This current, while propagating in the PL, becomes spin-polarized in the direction of the PL magnetization, and due to the *spin-transfer effect*[1,2] it transfers the spin angular momentum to the FL. For one direction of the current this transfer can destabilize the equilibrium orientation of the FL magnetization. In contrast, the natural magnetic damping caused mainly by the spin-electron interaction and characterized by the phenomenological damping constant $\alpha$ tries to bring the FL magnetization back to its equilibrium orientation. The dynamics of magnetization $\mathbf{M} = \mathbf{M}(t, \mathbf{r})$ in the FL under the action of spin-polarized current and natural dissipation is described by the Landau-Lifshits-Gilbert-Slonczewski (LLGS) equation[1]:

$$\frac{\partial \mathbf{M}}{\partial t} = \gamma \left[ \mathbf{H}_{\text{eff}} \times \mathbf{M} \right] + \frac{\alpha}{M_0} \left[ \mathbf{M} \times \frac{\partial \mathbf{M}}{\partial t} \right] + \frac{\sigma I}{M_0} f(r/R_c) \left[ \mathbf{M} \times \left[ \mathbf{M} \times \mathbf{p} \right] \right], \tag{1}$$

where $\gamma$ is the gyromagnetic ratio and $\mathbf{H}_{\text{eff}}$ is the effective magnetic field calculated as a variational derivative ($\mu_0 \mathbf{H}_{\text{eff}}(t,\mathbf{r}) = -\delta W / \delta \mathbf{M}$) of the magnetic energy $W$ of the system, which includes magnetostatic, exchange, and Zeeman contributions. The external bias field $\mathbf{H}_{ext}$ is considered to be constant in magnitude, but can have different orientations $\theta_{ext}$ with respect to the plane of the FL.

The second term in the right-hand side of Eq. (1) is the phenomenological magnetic damping torque written in the traditional Gilbert form, and $M_0 = |\mathbf{M}|$ is the saturation magnetization of the FL. The last term in the right-hand side of Eq. (1) is the Slonczewski spin-transfer torque[1,3] that is proportional to the bias current $I$. In Eq. (1) the function $f(r/R_c)$ characterizes the spatial distribution of the current across the area of the nanocontact (where $R_c$ is the nanocontact radius), and the coefficient $\sigma$ is related to the dimensionless spin-polarization efficiency $\varepsilon$ by the expression[3]:

$$\sigma = \frac{\varepsilon g \mu_B}{2 e M_0 S d_{FL}}. \qquad (2)$$

Here $g$ is the spectroscopic Landè factor, $\mu_B$ is the Bohr magneton, $e$ is the absolute value of the electron charge, $d_{FL}$ is the FL thickness, and $S = \pi R_c^2$ is the area of the nanocontact. The unit vector $\mathbf{p}$ defines the spin-polarization direction that coincides with the equilibrium direction of the magnetization of the PL.

In our approach, the LLGS equation (1) was numerically solved using our own 3D Finite-Differences Time-Domain (FD-TD) micromagnetic code that employs a fifth-order Runge-Kutta scheme (see Refs.12-14 and 18-21 for further details). The magnetodipolar field was computed using the Newell tensor,[22] while the exchange field was calculated assuming a six-neighbors interaction.

In our calculations we made several simplifying assumptions, similar to the ones used in our previous works.[13,14] The current-carrying nanocontact region was considered to be circular with the radius $R_c$. The current density distribution was assumed to be uniform within the contact region [$f(r/R_c) = 1$, if $r \leq R_c$] with an abrupt cut-off outside that region [$f(r/R_c) = 0$, otherwise]. Both the thickness and the saturation magnetization of the PL were assumed to be large enough to prevent any dynamics in this layer, so that the direction of the vector $\mathbf{p}$ is varied only through the variation of the external field angle and is independent of the applied current. We neglected the constant current-induced (Oersted) magnetic field and the magnetostatic coupling between the two ferromagnetic layers (FL and PL), since we believe that in the presence of a sufficiently large constant bias magnetic field $\mathbf{H}_{ext}$ these effects cannot qualitatively change the structure of spin wave modes excited in a nanocontact by the spin-polarized current. We, also, ignored the magnetocrystalline anisotropy in the FL, which is a usual assumption for magnetically soft Permalloy layers.

To reduce the computation time we, also, neglected the random fluctuations arising from the thermal noise. As explained in our previous works,[12,14] these fluctuations do not change the profile and the frequency of spin wave modes excited in a nanocontact. However, in a real laboratory experiment these thermal fluctuations of magnetization might create a finite level of magnetization oscillations that is necessary to excite the subcritically-unstable spin-wave "bullet" mode[14-15] when the bias current is progressively increased from zero to a finite value. In contrast, as it was demonstrated in Ref. 14, in a numerical simulation with no thermal effects the "bullet" mode can be excited only by starting from a large magnitude of the bias current (corresponding to a strongly nonlinear regime of magnetization oscillations) and gradually reducing the value of this current. Therefore, to observe all the possible spin wave modes in our noise-free computational framework, the numerical simulations for every orientation of the bias field were performed by progressively *increasing* and, then, *decreasing* the bias current magnitude (see Ref. 14 for details).

The parameters used in our present work to simulate the current-induced spin-wave dynamics in a Permalloy FL are: FL thickness $d_{FL} = 5$ nm, nanocontact radius $R_c = 20$ nm, spin-polarization efficiency $\varepsilon = 0.25$, saturation magnetization of the FL $\mu_0 M_0 = 0.7$ T, spectroscopic Landè factor $g = 2.0$, and exchange stiffness constant in the FL $A_{ex} = 1.4 \times 10^{-11}$ J/m. The magnitude $|\mathbf{H}_{ext}|$ of the external bias magnetic field was chosen to be $\mu_0 H_{ext} = 0.8$ T, and the field vector $\mathbf{H}_{ext}$ was assumed to lie always in the *xz* plane, while its direction was varied from the in-plane (along the *x* axis, $\theta_{ext} = 0$) to perpendicular-to-plane of the FL (along the *z* axis, $\theta_{ext} = 90°$), as it is shown in Fig. 1. This range of magnetization angle variation was explored using a step-size of 5°, which was reduced to 1° in the vicinity of singularities.

The parameters used to compute the equilibrium magnetic state of the $Co_{90}Fe_{10}$ PL are: thickness $d_{PL} = 20$ nm, saturation magnetization $\mu_0 P_0 = 1.88$ T, exchange stiffness constant $A_{ex\_PL} = 2.0 \times 10^{-11}$ J/m, and the cubic anisotropy constant $k_{ani} = 5.6 \times 10^4$ J/m$^3$.

As discussed in previous works,[11,13,14,20] the micromagnetic numerical simulations in the nanocontact geometry used in the experiments[7,8] (where the in-plane sizes of magnetic layers are much larger that the nanocontact radius) have an inherent difficulty related to prohibitively large computational times if the real lateral sizes of the magnetic nanostructures are used in the simulation. Thus, in our current numerical experiment we limited the computational region to be $L \times L \times d = 800$ nm $\times$ 800 nm $\times$ 5 nm and used a mesh of discretization cells having the sizes 4 nm $\times$ 4 nm $\times$ 5 nm. To reduce the spin wave reflections at the boundaries of the computational region we imposed *ad-hoc* absorbing boundary conditions by introducing a spatially-dependent dissipation function $\alpha = \alpha(r)$ (where $r$ is the distance from the center of the nanocontact), similar to the dissipation function used in our previous calculations.[11,13,14,20] In particular, in our current work the magnetic dissipation in the magnetic medium of the FL was assumed to be independent of the radial coordinate $r$ and equal to its physical value $\alpha_G$ (the dimensionless Gilbert damping constant) within a circular region of radius $R^* \gg R_c$, whereas outside this region the dissipation was assumed to increase linearly with coordinate $r$, and with a spatial rate $q$[13,14,20]:

$$\alpha(q,r) = \begin{cases} \alpha_G, & \text{if } r < R^* \\ \alpha_G(1 + q(r - R^*)), & \text{if } r > R^* \end{cases} \quad (3)$$

The value of the Gilbert damping constant was chosen to be $\alpha_G = 0.01$ (which is typical for good-quality Permalloy), while the other parameters of the dissipation function (3) $R^* = L/2 - 40$nm and $q = 100/(L/2 - R^*)$ were chosen empirically to minimize the reflection of the propagating wave in the numerical experiments, and, at the same time, to preserve the (physical) material properties in a computational area as good as we can.

An additional proof that our choice of the parameters in the dissipation function (3) is reasonable comes from the fact that the threshold of excitation of a linear spin wave mode, numerically calculated using the dissipation function (3), does not differ by more than 10 % from the corresponding threshold, analytically calculated using Eq. (13) in Ref. 3, for most orientations of the external bias field $\theta_{ext}$. Using a similar criterion, we have, also, numerically verified that the computational region having the in-plane sizes 800 nm × 800 nm is sufficiently large to give the reasonable quantitative values for all the calculated variables.

Another simplifying assumption used in our calculations was the assumption that the boundary conditions for both exchange and magnetostatic fields at the computational region boundaries are independent of the direction of the bias magnetic field, and have a following simple form

$$\left.\frac{\partial \mathbf{m}}{\partial n}\right|_{boundaries} = 0, \qquad (4)$$

where $n$ is the direction normal to the boundaries.

Although conditions (4) are not strictly rigorous, the only drawback created by them is the non-flat profile of the total effective field in the vicinity of the computational boundaries, which, in its turn, creates some additional spurious spin-wave reflections. Our previous investigations[14] have demonstrated, however, that the spin wave mode profiles calculated using the approximate boundary conditions (4) are sufficiently smooth, and the use of these conditions does not lead to any qualitative changes in the studied current-induced magnetization dynamics.

## RESULTS AND DISCUSSION

The results of our micromagnetic simulations of current-induced spin wave excitation for several different magnetization angles $\theta_{ext}$ are presented in Fig. 2, where we show the generated microwave frequency as function of the applied bias current. For each value of the magnetization angle $\theta_{ext}$, the simulations started from the equilibrium magnetization distribution and zero bias current. During the simulation, we slowly increased the bias current to a sufficiently large supercritical value and then reduced it back to zero. The solid arrows in Fig. 2 denote the branches observed during the increase of the bias current, while the dashed arrows denote the branches observed during the decrease of the current. Depending on the value of the magnetization angle $\theta_{ext}$, we have observed both hysteretic (see Fig. 2 (a-c)) and non-hysteretic (see Fig. 2 (d)) types of the spin wave excitation.

For the in-plane magnetized nanocontact $\theta_{ext} = 0$ (see Fig. 2 (a)), the spin-wave excitations at the branch corresponding to the increasing bias current start at a relatively large value of the bias current $I_{th}^{L} = 10$ mA equal to the threshold of excitation of a "quasi-linear" propagating spin wave mode, which was previously calculated analytically in Refs. 10 and 16. The frequency of this "quasi-linear" exchange–dominated spin wave mode is well above the ferromagnetic resonance (FMR) frequency of the FL. As it was demonstrated earlier,[14] this mode is nothing else but the analog of a linear propagating spin wave mode discovered by Slonczewski in Ref. 3 for the case of perpendicular magnetization. With the increase of the bias current the frequency of this Slonczewski's-like mode decreases slightly (demonstrating a "red" nonlinear frequency shift typical for the case of in-plane magnetization[17]) until the bias current reaches the upper critical value of about 11.5 mA, at which the propagating mode loses stability and transforms into a strongly nonlinear self-localized spin wave "bullet" mode,[10,14] having the frequency that is below the FMR frequency. With the further increase of the bias current the frequency of the bullet mode decreases slightly, but the structure of the mode does not experience any qualitative changes (see Fig. 2 (a)).

If now we begin to decrease the bias current, starting from the highly supercritical value (the branch corresponding to the decreasing current is denoted by the dashed arrows in Fig. 2 (a)), the "bullet" mode remains stable down to a rather small value of the bias current $I_{th}^B$ = 2.5 mA, which we assume to be the threshold current for "bullet" mode excitation. We believe that in real experiments, where thermal fluctuations and field inhomogeneities can provide a sufficient level of magnetization deviations from the equilibrium state, the "bullet" mode is excited at $I = I_{th}^B$, regardless of the direction of change of the bias current.

For small out-of-plane magnetization angles $\theta_{ext}$ the described above picture of spin wave excitation remains qualitatively unchanged. The linear excitation threshold $I_{th}^L$ stays almost constant, but the range of existence of the quasi-linear propagating mode increases due to the increase of the upper critical current. The threshold current of excitation of the "bullet" mode $I_{th}^B$ monotonically increases with the increase of the out-of-plane magnetization angle $\theta_{ext}$.

At a certain critical magnetization angle $\theta_{ext} = \theta_{cr}$ ($\theta_{cr}$ = 56° for the parameters of our simulations) the critical currents for the excitation of linear and "bullet" modes become equal, $I_{th}^L = I_{th}^B$ (see Fig. 2 (b)), and, in principle, either mode or both of them can be excited in a laboratory experiment. For larger magnetization angles (see Fig. 2 (c)) the linear excitation threshold $I_{th}^L$ is smaller than $I_{th}^B$, and the quasi-linear propagating spin wave mode should be excited first when the bias current is increased. This excitation with the further increase of the bias current is followed by an abrupt downward frequency jump corresponding to the transition from the quasi-linear propagating mode to a strongly nonlinear self-localized "bullet" mode.

When the magnetization angle increases further, $\theta_{ext} > \theta_{lin}$ ($\theta_{lin} = 62°$ in our case), only the linear propagating Slonczewski-like mode is excited. For such magnetization angles the dependence of the generated frequency on the bias current becomes non-hysteretic (see Fig. 2 (d)).

The results of the numerical simulations, shown in Fig. 2, demonstrate that, depending on the value of the external magnetization angle $\theta_{ext}$, there exist three qualitatively different scenarios of current-driven spin wave excitation in magnetic nanocontacts. For small magnetization angles $0° < \theta_{ext} < \theta_{cr}$ one observes excitation of a strongly nonlinear self-localized spin wave "bullet" with the frequency that is below the FMR frequency. For large magnetization angles $\theta_{ext} > \theta_{lin}$ only the linear propagating spin wave mode with frequency that is above the FMR frequency is excited. In the intermediate range of magnetization angles $\theta_{cr} < \theta_{ext} < \theta_{lin}$ the type of the excited mode depends on the value of bias current: for relatively small values of current the linear mode is excited, while for larger values of current the quasi-linear excited mode is abruptly transformed into a nonlinear "bullet" mode. At this critical point, one observes an abrupt downward jump of the generated frequency. These numerical results suggest that the frequency jumps observed experimentally in current-driven nanocontacts[8] can be explained, in some cases, by the above proposed mechanism of mode hopping (or mode transformation).

The numerically calculated threshold currents, corresponding to the excitation of a linear propagating spin-wave mode (dashed line) and nonlinear "bullet" mode (dash-dotted line) as function of the external bias field angle are shown in Fig. 3. The minimum threshold current, which should correspond to the excitation of a spin wave mode in a real laboratory experiment, is shown by the solid line. It is clear from Fig. 3, that the dependence of the threshold current on the magnetization angle is continuous, but has a typical kink at the critical magnetization angle $\theta_{ext} = \theta_{cr}$, where the transition from the "bullet" mode to a quasi-linear propagating mode takes place.

In contrast, the angular dependence of the spin wave frequency generated at the threshold (see main panel on Fig. 4) has a discontinuity at $\theta_{ext} = \theta_{cr}$. This frequency jump $\Delta f$, taking place because of the switching from "bullet" mode to a quasi-linear propagating mode, has the magnitude of the order of several GHz and depends, mainly, on the radius of the nanocontact $R_c$. Thus, for the nanocontact radius $R_c = 20$ nm this frequency jump has a rather large magnitude of $\Delta f_1 = 6$ GHz (see main panel on Fig. 4) and reduces to $\Delta f_2 = 3$ GHz for the nanocontact radius $R_c = 32$ nm (see inset in Fig. 4). At the same time, the critical angle $\theta_{cr}$, at which the transition between two excited modes occurs, is practically independent of the nanocontact radius.

It is interesting to compare our numerical results with the predictions of the weakly-nonlinear analytical theory.[16] In Fig. 5 we present angular dependences of the threshold current (see main panel in Fig. 5) and the spin wave frequency generated at the threshold (see inset in Fig. 5) calculated in the approximate analytical approach[16] for the same nanocontact parameters that were used in our numerical calculations (see Figs.3 and 4).

It is clear from the comparison of Fig. 5 with Figs. 3 and 4 that the weakly-nonlinear analytic theory gives threshold curves that are qualitatively similar to the corresponding curves obtained in full micromagnetic simulations. Surprisingly, even the quantitative values of the threshold current and the spin wave frequency generated at the threshold for a highly-nonlinear bullet mode are in good agreement with numerical results for sufficiently small magnetization angles $\theta_{ext} < 40°$, while the analogous comparison for a linear propagating Slonczewski mode gives a satisfactory agreement in the whole range $0° < \theta_{ext} < 90°$.

There are, however, some quantitative discrepancies between the analytic and numerical descriptions of the current-induced magnetization dynamics of a magnetic nanocontact. First of all, the critical magnetization angle at which the switching from a "bullet" mode to a quasi-linear mode takes place, is around $\theta_{cr} = 75°$ in a weakly-nonlinear theory[16] (see Fig. 5) and only $\theta_{cr} = 58°$ in the numerical calculation (see Figs. 3 and 4). The maximum angle at which the nonlinear "bullet" mode can exist is $\theta_{lin} = 77°$ in analytical approach and $\theta_{lin} = 62°$ in micromagnetic simulations. We attribute this discrepancy to the fact that in magnetic systems the *internal* magnetization angle (i.e., the angle determining the direction of the static equilibrium magnetization vector) depends on the amplitude of the excited spin waves.[17] Since the total length of the magnetization vector is constant, the excitation of spin waves with large precession angle reduces the static magnetization and, therefore, increases the internal magnetization angle. As a result, in the strongly excited magnetic system the nonlinear frequency shift coefficient $N$ vanishes at a smaller external angle $\theta_{ext}$ than predicted by the weakly-nonlinear theory,[16] where the expansion around the equilibrium magnetization direction was used and, respectively, nonlinear changes in the internal magnetization angle were ignored. An additional confirmation of this mechanism follows from the fact that in our numerical simulations the shift of the frequency of the linear propagating spin wave mode with the bias current changes from negative ("red" frequency shift, typical for systems with $N < 0$) to positive ("blue" frequency shift, characteristic for systems with $N > 0$) exactly at $\theta_{ext} = \theta_{lin} = 62°$.

At the same time, we would like to stress that the full quantitative description of the experimentally observed magnetization dynamics in current-driven magnetic nanocontacts is beyond the scope (and beyond the validity region) of the above described simplified numerical model. For instance, in the experiments,[7,8] apart from the fact that several different frequencies were simultaneously generated at certain magnitudes of the bias current, it was also observed that for a given external magnetization angle, the dependence of the generated frequency on the bias current for a particular mode can be non-monotonous[8] (i.e. a red-frequency shift with increasing current followed by a blue-frequency shift for larger current values). Such complicated behavior was not fully reproduced by the simple macrospin model,[8] or by the approximate theoretical approach of Ref. 16, or by the simplified micromagnetic modeling presented here.

We could only attribute this non-trivial magnetization dynamics to the contributions of the effective field neglected in all the above mentioned calculations. For instance, it has been demonstrated in excellent recent numerical simulations,[11,23,24] that the role played by the current-induced Oersted field can be qualitatively important for the current-induced magnetization dynamics in magnetic nanostructures. In particular, the numerical calculations performed in Ref. 11 in the presence of the Oersted field pointed out that for relatively large magnitudes of the bias current (that are sufficient to induce local magnetization reversal) the inhomogeneous spatial distribution of the Oersted field could create a non-monotonic dependence of the generated spin wave frequency on the bias current (see Fig. 5 in Ref. 11). Thus, a more sophisticated numerical micromagnetic modeling is necessary to describe all the features of laboratory experiments.

## CONCLUSION

In conclusion, we used a simplified micromagnetic model to study numerically the nature of the microwave spin-wave modes excited by spin-polarized current in a nanocontact geometry when the orientation of the external magnetic field is varied from in-plane to perpendicular-to-plane. To compensate for the lack of thermal noise in our model we did modeling with both increasing and decreasing bias current. This allowed us to investigate the regions of existence of subcritically-unstable[14,15] "bullet" spin wave modes. It was found that, with the increase of the out-of-plane magnetization angle, at a certain critical magnitude of this angle an abrupt jump in generated spin wave frequency occurs, and this jump is related to the hopping between the self-localized nonlinear spin wave "bullet" mode[10] and the quasi-linear propagating spin wave mode.[3] The numerically simulated spin wave dynamics is in qualitative agreement with the dynamic scenario predicted by the weakly-nonlinear analytical theory, but the critical magnetization angles corresponding to the mode hopping are substantially smaller in the numerical modeling than in the analytic theory. We believe that the analytically predicted and numerically confirmed scenario of mode hopping can explain some of the abrupt jumps in the generated microwave frequency observed in the laboratory experiments.[8] At the same time, it became clear that, although the simplified deterministic micromagnetic framework described above could partially explain the existence of multiple non-harmonically-related peaks,[8] it can not fully reproduce all the complexity of the experimentally observed microwave magnetization dynamics induced by spin-polarized current in nanocontact geometry[8] and a more sophisticated numerical model taking into account thermal fluctuations, dynamics of the pinned magnetic layer, and the Oersted field created by the bias electric current is necessary for the full description of experiments.

# ACKNOWLEDGEMENTS

The authors gratefully acknowledge Vito Puliafito for the software support. This work was supported in part by the MURI Grant No. W911NF-04-1-0247 from the Department of Defense of the USA, by Contract No. W56HZV-07-P-L612 from the U.S. Army TARDEC, RDECOM, by Grant No. ECCS-0653901 from the National Science Foundation of the USA, and by the Oakland University Foundation.

[*]email address: [consolo@ingegneria.unime.it](mailto:consolo@ingegneria.unime.it)

**FIGURE CAPTIONS**

**Fig. 1.** (Color online) Geometry of the point-contact device structure together with the coordinate system used in our simulations. The parameters shown in the figure are: $\mu_0 H_{ext}$ = 0.8 T, 0° ≤ $\theta_{ext}$ ≤ 90°, $R_c$ = 20 nm, $d_{FL}$ = 5 nm, $d_S$ = 5 nm, $d_{PL}$ = 20 nm, $L$ = 800 nm.

**Fig. 2.** (Color online) Dependence of the generated microwave frequency on the applied bias current for four external field (or magnetization) angles: (a) $\theta_{ext} < \theta_{cr}$, (b) $\theta_{ext} = \theta_{cr}$, (c) $\theta_{cr} < \theta_{ext} < \theta_{lin}$, (d) $\theta_{ext} > \theta_{lin}$. Arrows indicate the directions of current variation: solid line stands for the increasing current, dashed line - for the decreasing current. The dash-dotted vertical lines show the threshold currents corresponding to the excitation of quasi-linear $I_{th}^L$ and nonlinear "bullet" $I_{th}^B$ modes.

**Fig. 3.** (Color online) Dependence of the threshold current $I_{th}$ on the external field angle $\theta_{ext}$. The dashed and dash-dotted lines show the threshold currents $I_{th}^L$ and $I_{th}^B$, corresponding to the excitation of a quasi-linear mode and nonlinear "bullet" modes, respectively. The thinner vertical dotted lines show the critical angles: $\theta_{cr}$, at which the threshold currents for the excitation of the quasi-linear and nonlinear "bullet" modes are equal, and $\theta_{lin}$, above which the "bullet" mode does not exist.

**Fig. 4.** (Color online) Main panel: Dependence of the microwave precession frequency generated at the excitation threshold $f_{th}$ on the external field angle $\theta_{ext}$ for a nanocontact having the radius $R_c$ = 20 nm. The thinner vertical dotted lines show the critical angles $\theta_{cr}$ and $\theta_{lin}$. The dashed and dash-dotted lines represent the frequencies of the quasi-linear mode and the nonlinear "bullet" mode, respectively. An abrupt frequency jump $\Delta f_1$ takes place at the critical angle $\theta_{cr}$. Inset: Dependence

of the generated frequency $f_{th}$ on the magnetization angle for a larger nanocontact radius $R_c = 32$ nm, demonstrating a smaller frequency jump $\Delta f_2$.

**Fig. 5.** (Color online) Main panel: Theoretical dependence of the threshold current $I_{th}$ on the external field angle $\theta_{ext}$ calculated using the weakly-nonlinear formalism developed in Ref. 16. The dashed and dash-dotted lines represent the threshold current for the quasi-linear and nonlinear "bullet" modes, respectively. Thin dotted vertical lines indicate the critical angles $\theta_{cr}$ and $\theta_{lin}$. Inset shows the theoretical dependence of the frequency generated at the excitation threshold $f_{th}$ on the magnetization angle. Notations are the same as in the main panel.

**Fig. 1.**

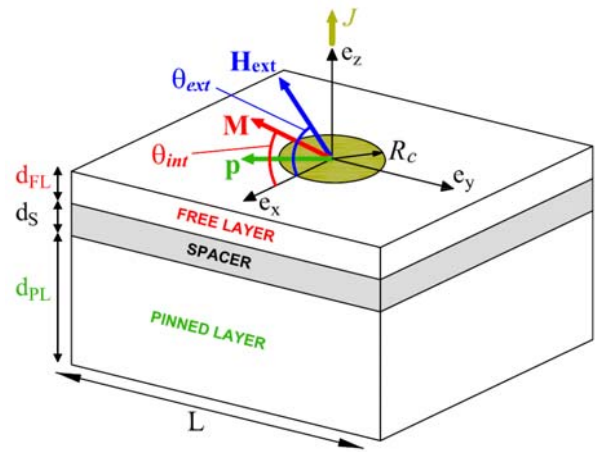

**Fig. 2.**

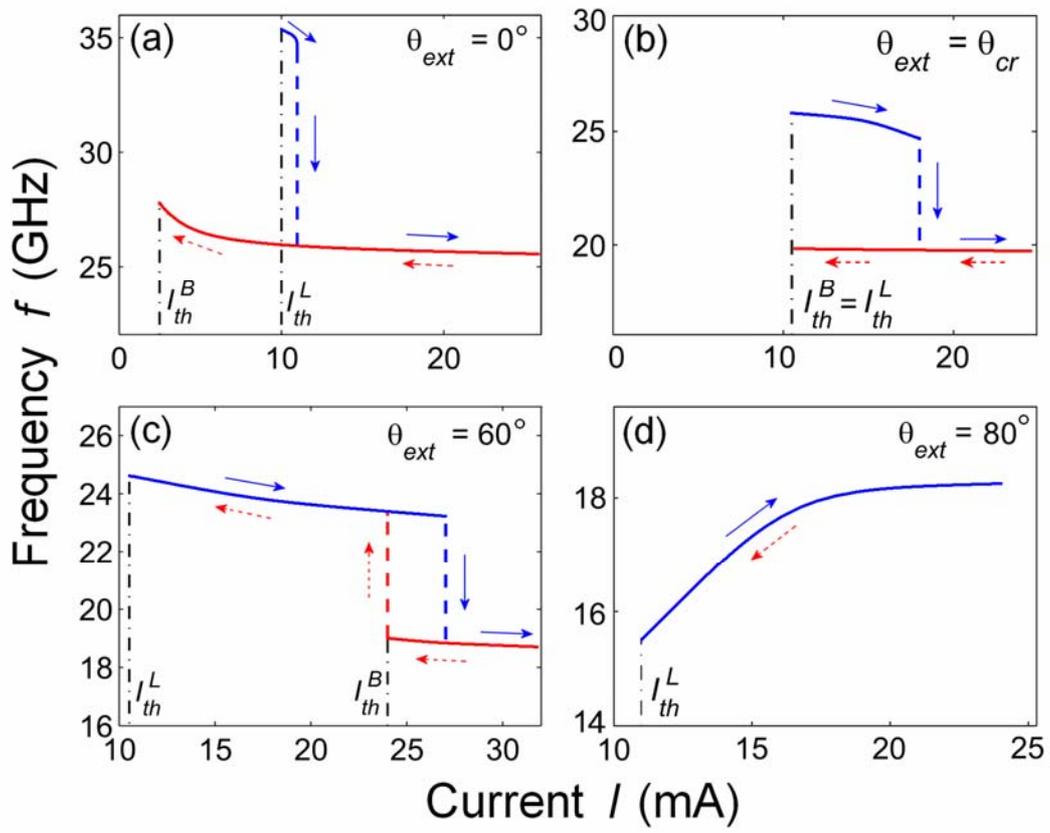

**Fig. 3.**

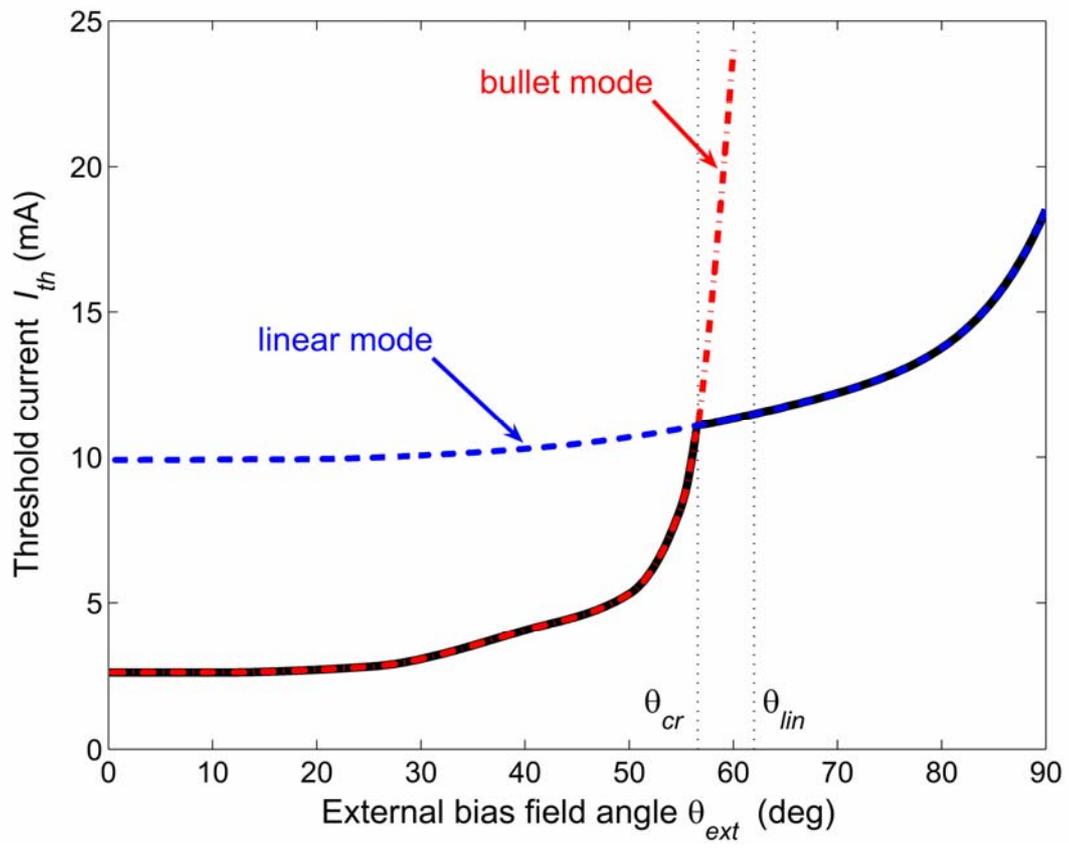

**Fig. 4.**

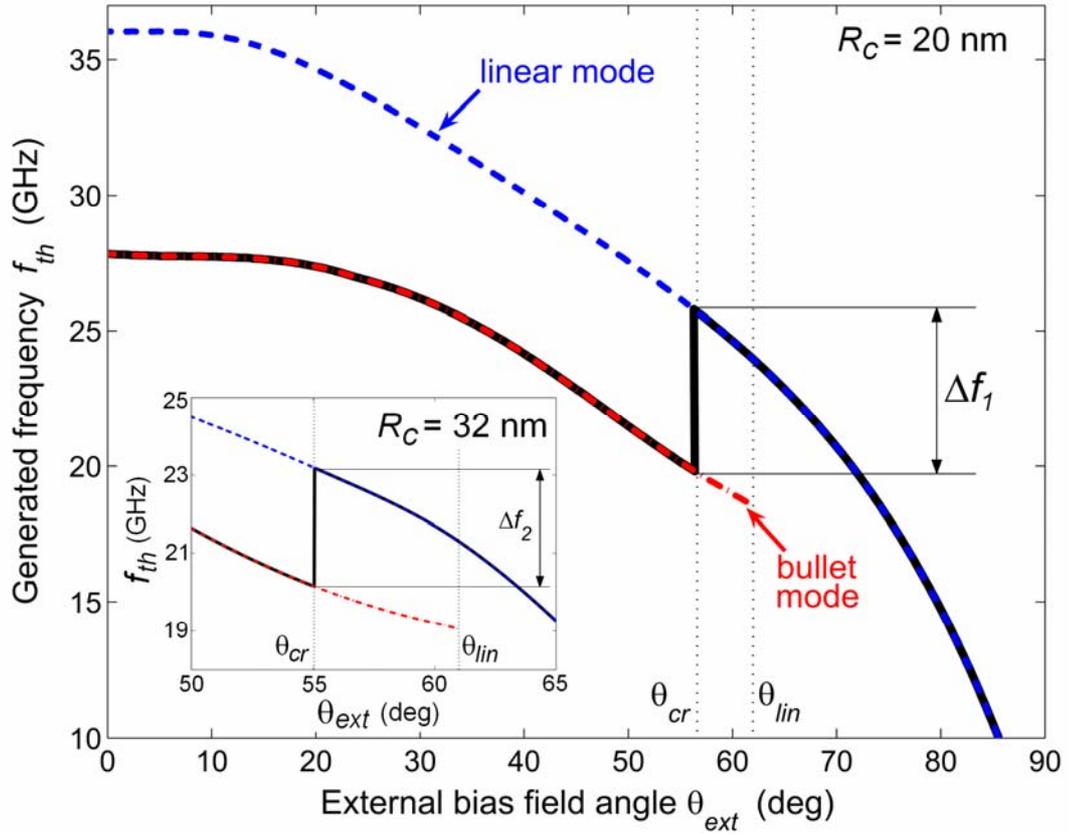

**Fig. 5.**

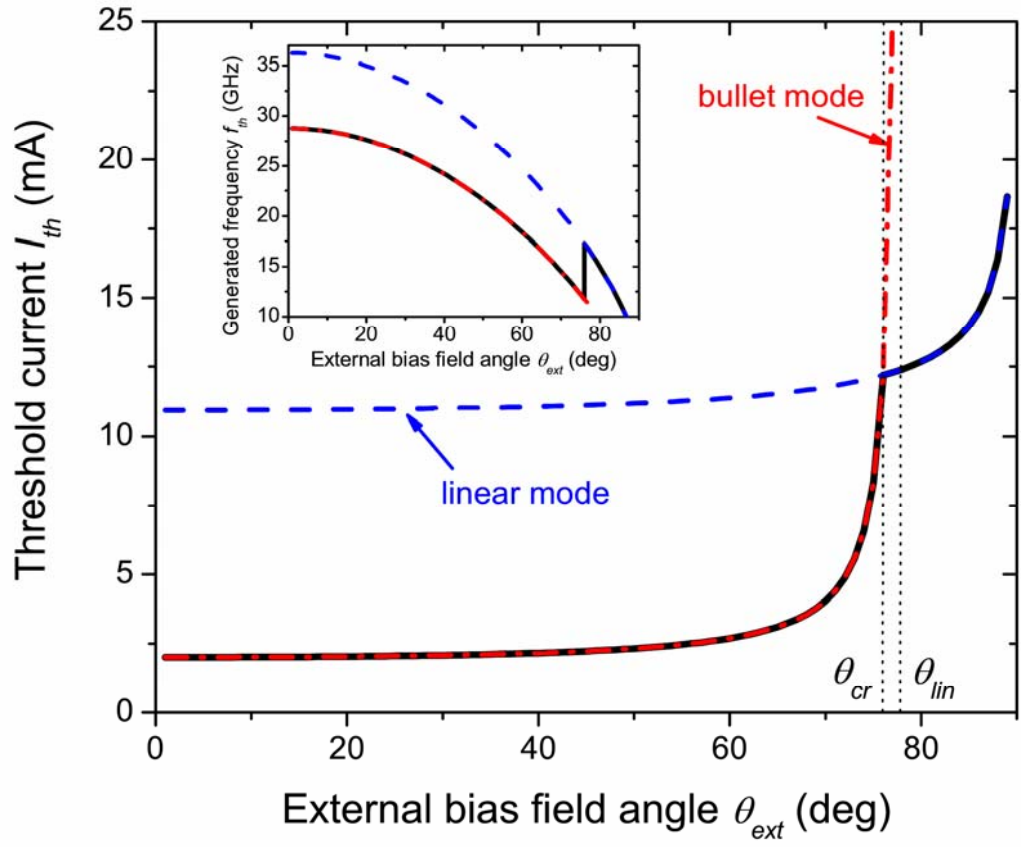